\documentclass[onecolumn,showpacs,superscriptaddress,pra]{revtex4}%
\usepackage{amsmath}
\usepackage{amsfonts}
\usepackage{amssymb}
\usepackage{graphicx}

\begin{document}

\title{Information Dissipation in Random Quantum Networks}

\author{Umer Farooq}
\address{School of Science and Technology, Physics Division, University of Camerino,
62032 Camerino, Italy\\
 and INFN-Sezione di Perugia, Via A. Pascoli, I-06123 Perugia, Italy}
 \email{umer.farooq@unicam.it}

 \author{Stefano Mancini}
 \address{School of Science and Technology, Physics Division, University of Camerino,
62032 Camerino, Italy\\
 and INFN-Sezione di Perugia, Via A. Pascoli, I-06123 Perugia, Italy}
\email{stefano.mancini@unicam.it}

\begin{abstract}
     We study the information dynamics in a network of spin-$1/2$ particles when edges representing $XY$ interactions are randomly added to a disconnected graph accordingly to a probability distribution characterized by a ``weighting" parameter. In this way we model dissipation of information initially localized in single or two qubits all over the network. We then show the dependence of this phenomenon from weighting parameter and size of the network.
\end{abstract}

\pacs{03.65.Yz, 02.50.Ey, 03.67.-a}

\maketitle

\section{Introduction}

A quantum network consists of spin-$1/2$ particles (qubits) attached to the nodes of a graph and interacting according its adjacency matrix (edges distribution) \cite{bose07}. This quantum model allows to accomplish information manipulation tasks that are impossible in the classical realm and it can be regarded as prototype model for a future quantum internet \cite{qi,go,pm,qnt}.  Within such a quantum network model information processing is usually described by assuming perfect control of the underlying graph. However this is not much realistic since randomness is often present and it leads to decoherence effects \cite{petr}. In contrast, the conservation of coherence is essential for any quantum information process \cite{bez}, hence there is a persistent interest in decoherence of pure states. To describe it the master equation approach is widely used. This employes Lindblad superoperators to phenomenologically add the decoherence process to Hamiltonian dynamics \cite{petr}. Recently random matrices, originally introduced by Wigner \cite{wig} and then used to study spectral statistics of dynamical systems (both  classical and quantum) \cite{mehta}, have been also employed to describe decoherence processes \cite{j7,j11,prof}.

Here we proceed along a similar line by assuming randomness of the graph (adjacency matrix) underlying the quantum network. We study the simplest task, namely information storage (in single and two qubits), when the graph underlying a network randomly changes in time. More specifically we consider the evolution of a quantum state in single and two qubits on nodes of initially disconnected graph when edges, representing $XY$ interactions, are randomly added.
To this end we employ a standard model for random graphs, that of Gilbert characterized by
a binomial distribution with a weighting parameter \cite{Gilbert}. By means of tools like fidelity, concurrence and von Neumann entropy \cite{bez} we see how information spreads all over the quantum network, i.e. dissipated. We then show the dependence of this quantum information dissipation phenomenon from weighting parameter and size of the network.

The article is organized as follows. In Sect. $2$ we introduce the random network model relying on $XY$-type
spin-spin interaction confining our attention to the single excitation subspace. In Sect. $3$ we discuss information dissipation from a single qubit, while in Sect. $4$ we investigate the two qubit entanglement dissipation. Conclusions are drawn in Sect. $5$.


\section{Networks Model}

Let $G = (V, E)$ be a simple undirected graph (that is, without loops or parallel edges), with set of vertices $V(G)$ and set of edges $E(G)$. We take $V(G)=\{1, . . . , N\}$, being $N$ the number of nodes. The adjacency matrix of $G$ is denoted by $A(G)$ and defined by $[A(G)]_{ij} = 1$, if $i$ $j$ $\in E(G)$; $[A(G)]_{ij} = 0 $ if $i j \notin E(G)$. The adjacency matrix is a useful tool to describe a network of $N$ spin-1/2 quantum particles. The particles are usually attached to the vertices of $G$, while the edges of $G$ represent their allowed couplings. For the $XY$ interaction model (isotropic Heisenberg model), ${ij}\in E(G)$ means that the particles $i$ and $j$ interact by the Hamiltonian $[H(G)]_{ij}= (X_{i}X_{j}+Y_{i}Y_{j})$, where $X_{i}$ and $Y_{i}$  are the Pauli operators of the $i$-th particle. The Hamiltonian of the whole network then reads
\begin{equation}\label{H}
H_{XY}(G)=\sum_{i\neq j}[A(G)]_{ij}(X_{i}X_{j}+Y_{i}Y_{j}).
\end{equation}
Here, the number of edges  $n \in [0,n_{max}], $ where $n_{max}=N(N-1)/2$, will be chosen randomly.
A standard way for doing that is given by the Gilbert's random graphs model
characterized by a binomial probability distribution
 \begin{equation} \label{pn}
 p(n)=
 \left(\begin{array}{c}
 n_{max}\\
 n
 \end{array}\right)
 \xi^n\, (1-\xi)^{n_{max}-n},
 \end{equation}
 where {\scriptsize $\left(\begin{array}{c}
 n_{max}\\
 n
 \end{array}\right)$} is the binomial coefficient and $\xi$ is a parameter such that $0\le \xi \le 1$ \cite{Gilbert}.

 Notice that all configurations with a given number $n$ of edges are considered equally probable.
 The parameter $\xi$ can be thought as a weighting parameter; by increasing its value  the model becomes more and more likely to include graphs with more edges and less and less likely to include graphs with fewer edges. The case of $\xi=0.5$ corresponds to the situation where all graphs are chosen with equal probability.

 This can be turned into the adjacency matrix $A(G)$ in the following way
\begin{eqnarray}\label{a}
Pr[A_{ii}=1]&=&0, \nonumber\\
Pr[A_{ii}=0]&=&1, \nonumber\\
Pr[A_{ij}=1]&=& \xi,\quad\quad \text{for}\quad i > j, \nonumber\\
Pr[A_{ij}=0]&=& 1-\xi, \quad \text{for}\quad i > j, \nonumber\\
Pr[A_{ij}]&=&Pr[A_{ji}], \quad \text{for}\quad i < j.
\end{eqnarray}
Thus $A(G)$ is symmetric and the average number of edges (number of ones in the upper or lower triangular part of the adjacency matrix) will results $n_{max}\xi$ accordingly to
the distribution $p(n)$ of Eq.\eqref{pn}.

The Hamiltonian $H_{XY}$ in Eq.\eqref{H}, constructed according to the adjacency matrix $A(G)$ of Eq.\eqref{a}, results an operator on the complex Hilbert space $ \mathcal{H}\simeq (\mathbb{C}^2)^{N}$ of $N$
 spin particles and can be represented by a $2^{N} \times 2^{N}$ matrix. However, since such Hamiltonian conserves the number of excitations, the effective Hilbert space where the dynamics take place will be determined by the initial number of excitations. Below we will restrict our attention to cases with network initially having at maximum one excitation.
Hence, we are legitimate to consider the subspace $\mathcal{K}$ $\simeq \mathbb{C}^{N}$ spanned by
\begin{equation}
\{|1\rangle_1|0\rangle_2\ldots |0\rangle_N, |0\rangle_1|1\rangle_2\ldots |0\rangle_N,\ldots, |0\rangle_1|0\rangle_2\ldots |1\rangle_N\}.
\end{equation}
We will use the short hand notation
\begin{equation}
| j \rangle \equiv|0\rangle_1\ldots |0\rangle_{j-1} |1\rangle_j |0\rangle_{j+1}\ldots |0\rangle_N, \quad j=1,\ldots,N.
\end{equation}
The representation of the Hamiltonian $H_{XY}$  in the basis $\{\vert j\rangle\}_{j=1}^{N}$ exactly coincides with the adjacency matrix $A(G)$.
Below we will also consider the vector  $|j=0\rangle$, which corresponds to the state of the network with no excitations. Hence the space of states becomes $\mathcal{K}\bigoplus\{\vert 0\rangle\}$.

In order to describe the evolution of an initial network state $\vert \psi (0)\rangle\in \mathcal{K}\bigoplus\{\vert 0\rangle\}$, we will consider time discretized by steps of width $\Delta t$ labeled by a natural number $k\in\mathbb{N}$.
 At each time step we randomly generate the adjacency matrix according to Eq.\eqref{a}, i.e. the Hamiltonian $H_{XY}$ which then becomes dependent on the time step label  $k$. Hence within the $k$-th time step we have the unitary evolution given by
 \begin{equation}\label{u}
 U(t_{k},t_{k-1})=e^{-iH_{XY}(k)\Delta t},
 \end{equation}
 with $t_k=k\Delta t$.
 Then, let us consider the composition of such unitaries as follows
 \begin{equation}\label{evol}
{\cal U}(t_k)=U(t_k,t_{k-1}) \ldots U(t_2,t_1) U(t_1,t_0).
 \end{equation}
 It results, thanks to Eqs.\eqref{u} and \eqref{H}, that
 \begin{eqnarray}
{\cal U}(t_k)|0\rangle&=&|0\rangle,\\
{\cal U}(t_k)|i\rangle&=&\sum_{j=1}^Nu_{ij}(t_k)|j\rangle,
\end{eqnarray}
with $u_{ij}$ entries of a unitary matrix such that $u_{ij}(t_0)\equiv u_{ij}(0)=\delta_{ij}$.
Writing
\begin{equation}
\label{psi}
\vert \psi(0)\rangle=\sum^{N}_{i=0} c_{i} \vert i \rangle.
\end{equation}
with $c_{i}$ $\in$ $\mathbb{C}$ such that $\sum_{i=0}^N |c_{i}|^{2}=1$,
we will have
 \begin{equation}
\label{psitk}
\vert \psi(t_k)\rangle=c_0|0\rangle+\sum^{N}_{i,j=1} c_{i}u_{ij}(t_k) \vert j \rangle.
\end{equation}
This gives one ``realization" of a stochastic state evolution (say a quantum trajectory).
The average state of the network will be a density operator
\begin{equation}\label{e}
\varrho(t_k)=\mathbb{E}(|\psi(t_k)\rangle \langle\psi(t_k)\vert),
\end{equation}
where $\mathbb{E}$ represents the ensemble average over all possible realizations according to the probability distribution of Eq.\eqref{a}.


\section{Information Dissipation}

In this section, we will see how the information interchange between the spin particles and spread in the entire network, thus dissipating from one node to the others.
Let us consider the single spin particle labeled by $1$ in a generic superposition of ground and excited states and all the other spin particles in ground state. Then the initial state of the network reads
 \begin{equation}\label{inistate}
 \vert \psi(0) \rangle=\cos\frac{\theta}{2} \vert 0 \rangle +e^{i\phi}\sin\frac{\theta}{2}\vert 1 \rangle,
 \end{equation}
with $\theta\in[0,\pi]$ and $\phi\in[0,2\pi]$.
The evolution will be described by Eq.\eqref{psitk} where now
\begin{eqnarray}
c_0&=&\cos\frac{\theta}{2}, \\
c_1&=&e^{i\phi}\sin\frac{\theta}{2},\\
c_{i>1}&=&0.
\end{eqnarray}
Being interested in the information dissipation of  spin particle 1, we trace Eq.\eqref{e} over-all other particles
\begin{equation}\label{tr}
\varrho_{1}(t_k)=Tr_{N,N-1,N-2,É3,2}\left[\varrho(t_k)\right].
\end{equation}
It results
\begin{eqnarray}
\varrho_{1}(t_k)&=&\left[\cos^2\frac{\theta}{2}+\sin^2\frac{\theta}{2}\left(1-\mathbb{E}(|u_{11}(t_k)|^2)\right)\right]
|0\rangle_1\langle 0|\nonumber\\
&+&\cos\frac{\theta}{2}\sin\frac{\theta}{2} e^{-i\phi}\mathbb{E}(u_{11}^*(t_k)) |0\rangle_1\langle 1|\nonumber\\
&+&\cos\frac{\theta}{2}\sin\frac{\theta}{2} e^{i\phi}\mathbb{E}(u_{11}(t_k)) |1\rangle_1\langle 0|\nonumber\\
&+&\sin^2\frac{\theta}{2} \mathbb{E}(|u_{11}(t_k)|^2) |1\rangle_1\langle 1|.
\end{eqnarray}
Since fidelity is a standard tool used to study how different are two states \cite{bez},
we consider it between $ \varrho_{1}(t_k)$ and the initial state of Eq.\eqref{inistate}
 \begin{eqnarray}\label{f}
{\cal F}(t_k;\theta,\phi)&=&\Bigg|\left(\cos\frac{\theta}{2}
{}_{1}\langle 0\vert+e^{-i\phi}\sin\frac{\theta}{2}
{}_{1}\langle 1 \vert \right)  \varrho_{1}(t_k) \nonumber\\
&&\hspace{0.3cm}
\left(\cos\frac{\theta}{2}
|0\rangle_1+e^{i\phi}\sin\frac{\theta}{2}
|1\rangle_{1} \right) \Bigg|.
 \end{eqnarray}
Then, we determine the average (over all possible initial pure states)  fidelity as
 \begin{equation}\label{F}
 \overline{\cal F}(t_k)=\frac{1}{4\pi}\int_{0}^{2\pi}d\phi\int_{0}^{\pi} d\theta \sin\theta\; {\cal F}(t_k;\theta,\phi).
 \end{equation}
 Using Eqs.\eqref{psitk}, \eqref{e} into Eq.\eqref{f} and performing the integrals of Eq.\eqref{F} we get
 \begin{eqnarray}
\overline{\cal F}(t_k)=\frac{1}{2}+\frac{1}{3}\Re\left[\mathbb{E}\left(u_{11}(t_k)\right)\right]+\frac{1}{6}\mathbb{E}\left(|u_{11}(t_k)|^2\right).
\end{eqnarray}
For numerical calculations we have performed the ensemble average over $400$ realizations. We show in Fig.$1$ the average fidelity versus time $t_k$, for a network of $32$ spin particles at different values of $\xi$. For $\xi$ close to zero the dynamics results frozen with $\overline{\cal F}=1$ at any time in agreement with the fact that the most likely graph is the fully disconnected one.
By increasing the value of $\xi$ it can be clearly seen a decreasing behavior of the average fidelity versus time. Nevertheless oscillations appear and they tend to have a less damped amplitude. Finally, for $\xi$ close to $1$ we have a purely oscillatory behavior (see inset). This comes from  the fact that in this situation the most likely graph is the fully connected one, which causes a periodical back flow of information onto spin particle (qubit) $1$.
 \begin{figure}[htb]
\begin{center}
  \includegraphics[scale=0.2]{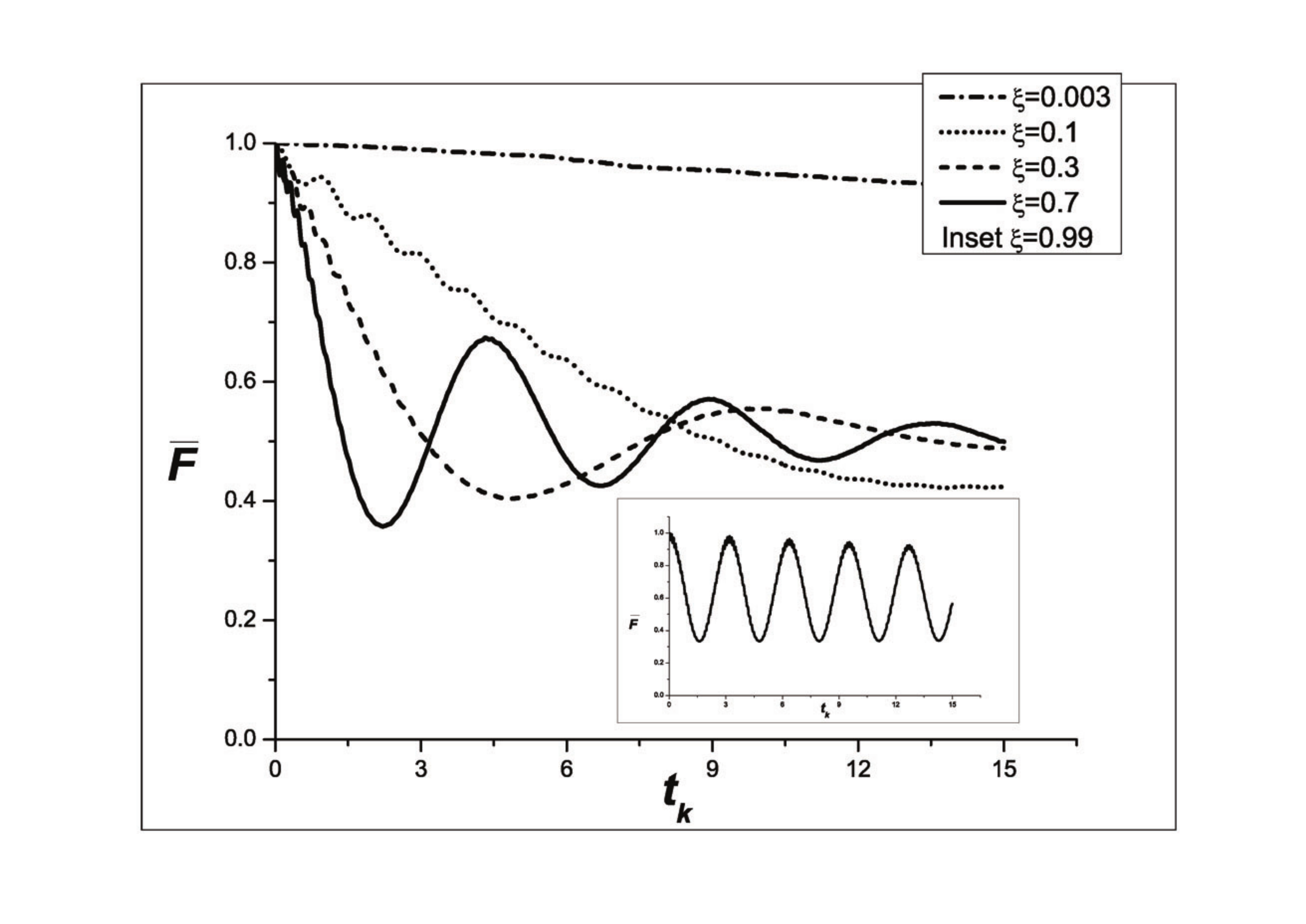}
  \caption{ The average fidelity $\overline{\cal F}$ versus (dimensionless) time $t_k$ for $N=32$ and $\Delta t= 0.015$.
  }
  \label{aaa}
\end{center}
\end{figure}

An effect similar to that of the increase of $\xi$,
namely the appearance of oscillations of more and more marked amplitudes,
becomes evident by reducing the size of the network (the number $N$) while keeping fixed the value of $\xi$ as shown in in Fig.$2$.
\begin{figure}[htb]
\begin{center}
  \includegraphics[scale=0.2]{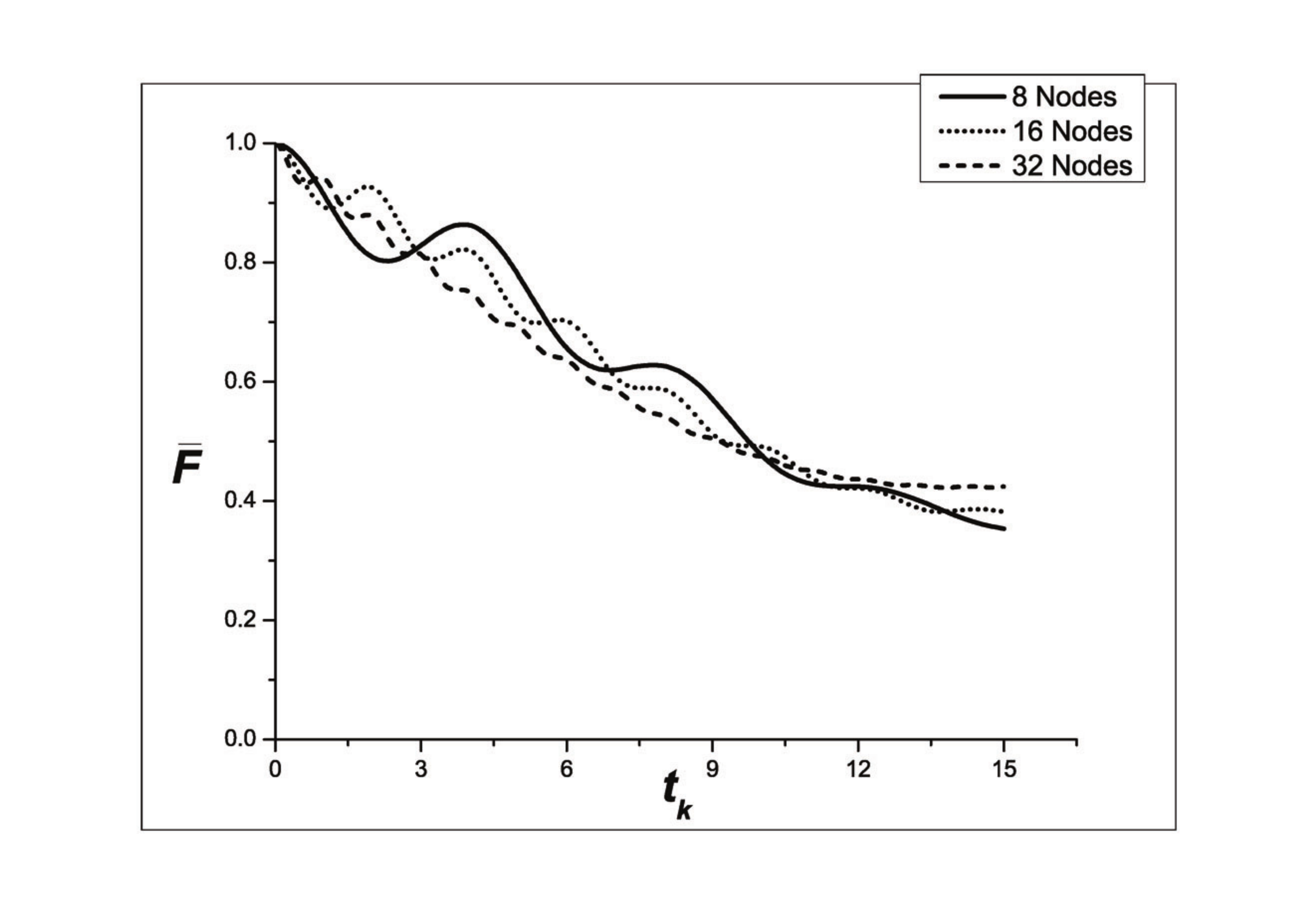}
  \caption{ The average fidelity $\overline{\cal F}$ versus (dimensionless) time $t_k$ for $\xi=0.3$ and $\Delta t=0.015$.
  }
  \label{bbb}
\end{center}
\end{figure}

The initial pure state of Eq.\eqref{inistate} for qubit 1 evolves into a mixed state (see Eq.\eqref{tr}).
Its degree of mixedness can be evaluated by means of the linearized von Neumann entropy
 \begin{equation}
{\cal S}^L_1(t_k;\theta,\phi)=1-Tr\left[\left(\varrho_{1}(t_k) \right)^2\right],
 \end{equation}
 ranging from zero for pure state to $1/2$ for a maximally mixed state $\mathbb{I}_1/2$ (being $\mathbb{I}_1$
 the identity operateor for the qubit $1$.
Likewise Eq.\eqref{F} the average (over all possible initial pure states) linear entropy reads
  \begin{equation}\label{S}
 \overline{{\cal S}^L_1}(t_k)=\frac{1}{4\pi}\int_{0}^{2\pi}d\phi\int_{0}^{\pi} d\theta \sin\theta\; {\cal S}(t_k;\theta,\phi).
 \end{equation}
Using Eqs.\eqref{psitk}, \eqref{e} into Eq.\eqref{f} and performing the integrals of Eq.\eqref{S} we finally get
\begin{eqnarray}
\overline{{\cal S}_1^L}(t_k)&=&\mathbb{E}\left(|u_{11}(t_k)|^2\right)-\frac{1}{3}\Big|\mathbb{E}\left(u_{11}(t_k)\right)\Big|^2 -\frac{2}{3}\left[\mathbb{E}\left(|u_{11}(t_k)|^2\right)\right]^2.
\end{eqnarray}
\begin{figure}[htb]
\begin{center}
  \includegraphics[scale=0.2]{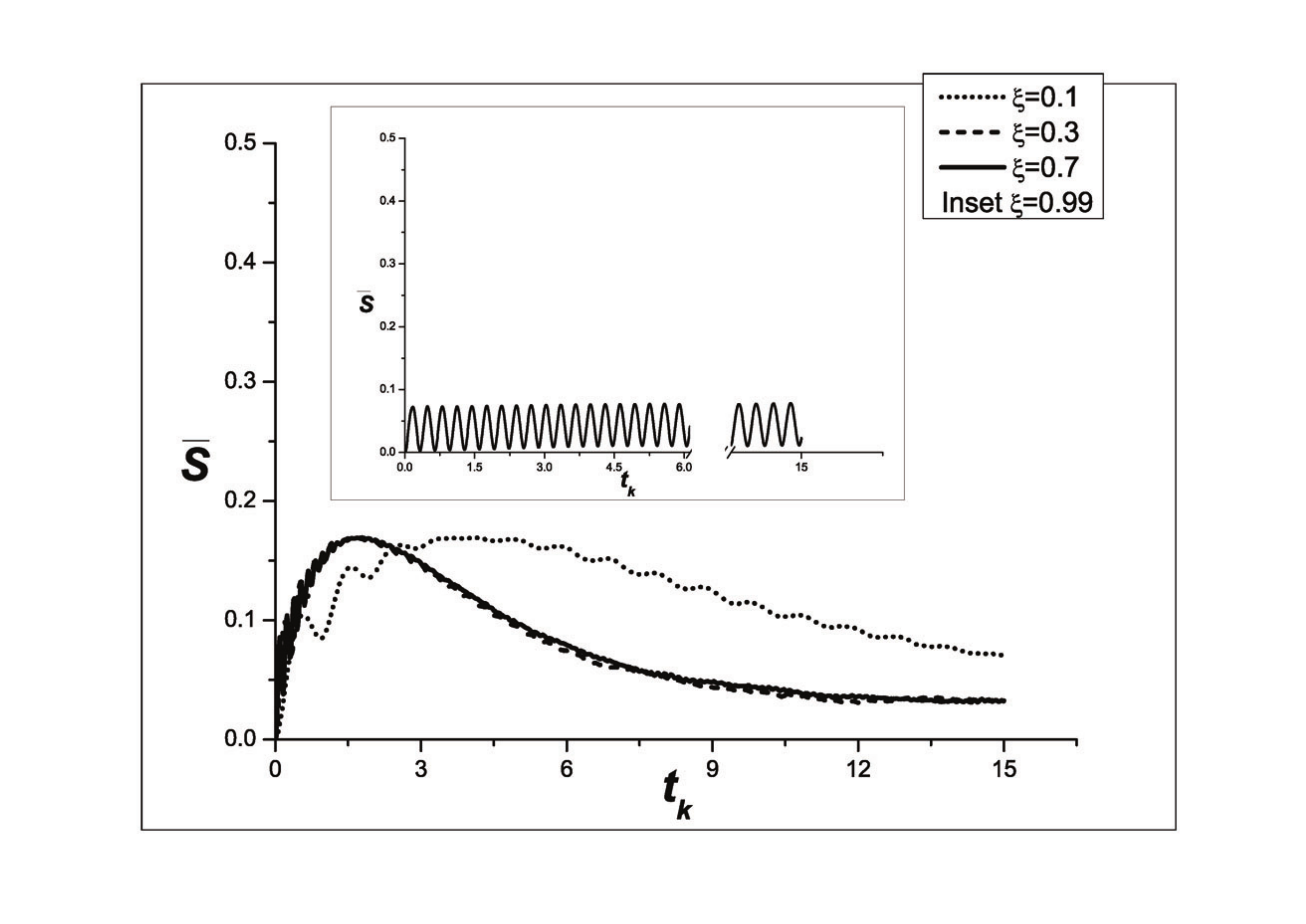}
  \caption{ The average linear entropy $\overline{{\cal S}_1^L}(t_k)$ versus (dimesnionless) time $t_k$ for $N=32$ and $\Delta t=0.015$.
  }
  \label{ccc}
\end{center}
\end{figure}

We show in Fig.$3$ the average linear entropy $\overline{{\cal S}_1^L}(t_k)$ versus time $t_k$ for a network of $32$ spin particles.
We can see that for small values of $\xi$ the entropy rises up more slowly and persists to an higher value for longer time. On the contrary, for larger values of $\xi$, due to higher connectivity, the dissipation (mixing) is faster, but the chances
to have back flow of information is higher. Thus after an initial bump the entropy decreases faster. Finally for $\xi\to1$ also the entropy becomes purely oscillatory in time with minima approaching zero, meaning that the state is cyclically purified (see inset).
By comparison of the insets of Figs.\ref{aaa} and \ref{ccc} we see that the oscillations of entropy are much more frequent. This means that the dissipated information most of the times bounces back in a state different from the initial one. Clearly maxima of fidelity occur in correspondence of some minima of entropy.


\section{Entanglement Dissipation}

In this section, we will see how the entanglement of an initially entangled pair of particles will spread all over the network.
We then consider to initially have particles labeled by $1$ and $2$ in a Bell state and all the other in the ground state. That is, the initial state of the network is
\begin{align}
|\psi(0)\rangle=\frac{1}{\sqrt{2}}( \vert1\rangle+\vert2\rangle).
\label{inistateent}
\end{align}
Notice that this state is a superposition of states containing a single excitation.
Hence, we can still use Eq.\eqref{psitk} where now
\begin{eqnarray}
c_0&=&0, \\
c_1&=&c_2=\frac{1}{\sqrt{2}},\\
c_{i>2}&=&0,
\end{eqnarray}
and then consider the density operator of Eq.\eqref{e}.
We are interested in the entanglement evolution
between particles $1$ and $2$ whose state is obtained by
\begin{equation}\label{trc}
\varrho_{12}(t_k)=Tr_{N,N-1,....4,3}(\varrho(t_k)).
\end{equation}
It results
\begin{align}
\varrho_{12}(t_k)&=\frac{1}{2}\left[2-\mathbb{E}\left(|u_{11}(t_k)+u_{21}(t_k)|^2\right)
-\mathbb{E}\left(|u_{12}(t_k)+u_{22}(t_k)|^2\right)\right]
|00\rangle_{12}\langle 00|\nonumber\\
&+\frac{1}{2}\mathbb{E}\left(|u_{11}(t_k)+u_{21}(t_k)|^2 \right)|10\rangle_{12}\langle 10|\nonumber\\
&+\frac{1}{2}\mathbb{E}\left(|u_{12}(t_k)+u_{22}(t_k)|^2\right) |01\rangle_{12}\langle 01|\nonumber\\
&+\frac{1}{2}\mathbb{E}\left((u_{11}(t_k)+u_{21}(t_k))(u_{12}^*(t_k)+u_{22}^*(t_k))\right) |10\rangle_{12}\langle 01|\nonumber\\
&+\frac{1}{2}\mathbb{E}\left((u_{11}^*(t_k)+u_{21}^*(t_k))(u_{12}(t_k)+u_{22}(t_k))\right)
|01\rangle_{12}\langle 10|.
\label{rho12}
\end{align}
A useful tool to measure the amount of entanglement is the concurrence defined as \cite{niel}
\begin{equation}\label{cnn}
{\cal C}_{12}(t_k)=\max\left(0,\sqrt{\lambda_{1}} - \sqrt{\lambda_{2}} - \sqrt{\lambda_{3}} - \sqrt{\lambda_{4}}\right),
\end{equation}
where $\lambda_{1}, \lambda_{2}, \lambda_{3} $ and $\lambda_{4}$ are the eigenvalues of $\varrho_{12}(t_k)\breve{\varrho}_{12}(t_k)$ in the decreasing order. Here
\begin{equation}
\breve{\varrho}_{12}(t_k)=(\mathbb{I}_1\otimes Y_2)\rho_{12}^*(t_k)(\mathbb{I}_1\otimes Y_2),
\end{equation}
where $\varrho_{12}^*(t_k)$  is the complex conjugate of $\varrho_{12}(t_k)$.

\begin{figure}[htb]
\begin{center}
  \includegraphics[scale=0.2]{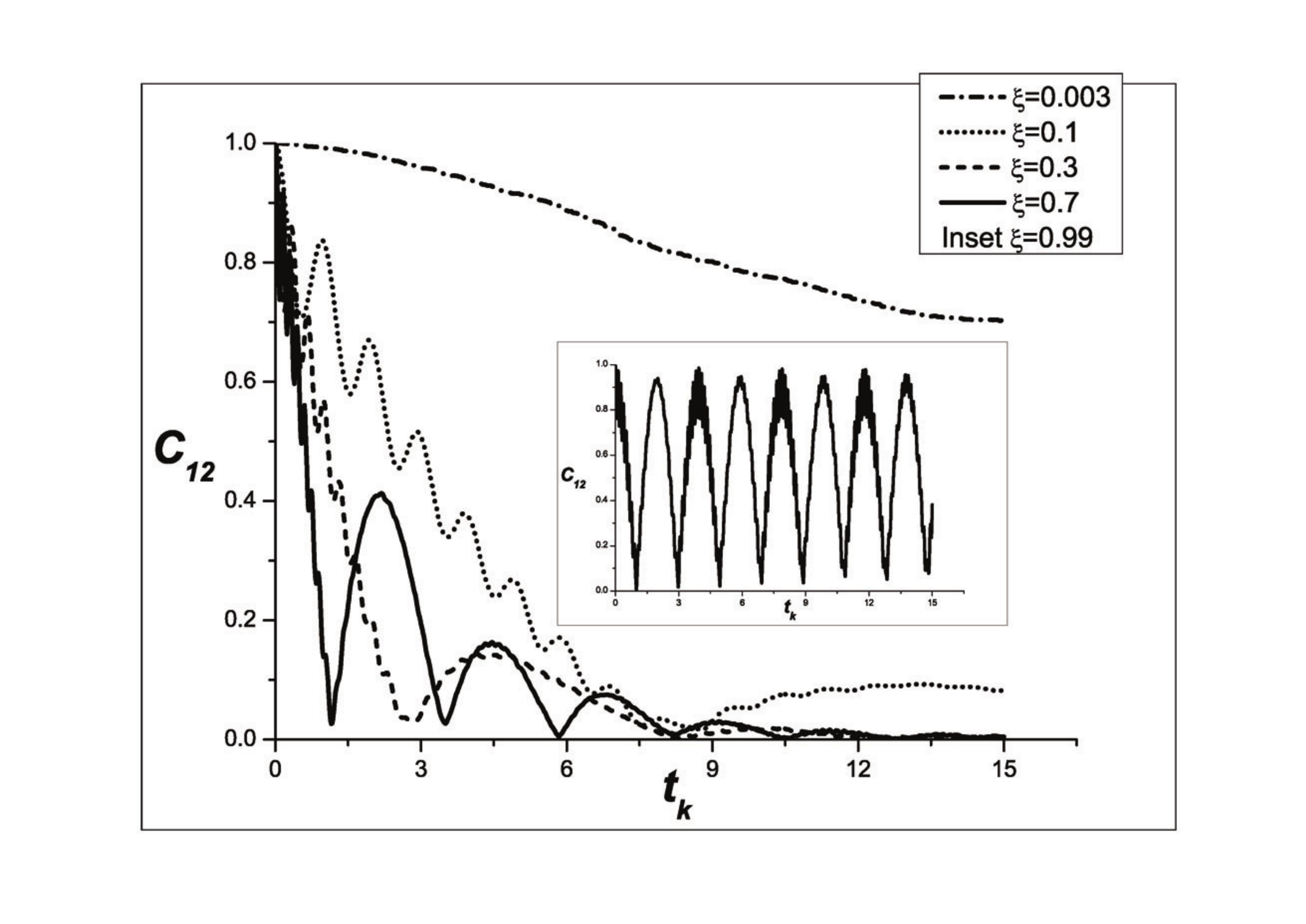}
  \caption{The concurrence ${\cal C}$ versus (dimensionless) time $t_k$ for $N=32$ and $\Delta t=0.015$.
  }
 \label{ddd}
\end{center}
\end{figure}

Also in this case for numerical calculations we have performed the ensemble average over $400$ realizations.  We show in Fig.\ref{ddd} the concurrence ${\cal C}_{12}(\varrho)$ versus time $t_k$, for a network of $32$ spin particles at different values of $\xi$.  It can be seen a decaying behavior of ${\cal C}_{12}(\varrho)$ versus time.
Pronounced oscillations appear by increasing the value of $\xi$ until the behavior becomes purely oscillator for $\xi$ close to 1 (similarly to what happen for the average fidelity).

In addition we have considered the evolution of the state
between particles $1$ and $3$
\begin{equation}
\varrho_{13}(t_k)=Tr_{N,N-1,\ldots,4,2}(\varrho(t_k)),
\end{equation}
resulting as
\begin{align}
\varrho_{13}(t_k)&=\frac{1}{2}\left[2-\mathbb{E}\left(|u_{11}(t_k)+u_{21}(t_k)|^2\right)
-\mathbb{E}\left(|u_{13}(t_k)+u_{23}(t_k)|^2\right)\right]
|00\rangle_{13}\langle 00|\nonumber\\
&+\frac{1}{2}\mathbb{E}\left(|u_{11}(t_k)+u_{21}(t_k)|^2 \right)|10\rangle_{13}\langle 10|\nonumber\\
&+\frac{1}{2}\mathbb{E}\left(|u_{13}(t_k)+u_{23}(t_k)|^2\right) |01\rangle_{13}\langle 01|\nonumber\\
&+\frac{1}{2}\mathbb{E}\left((u_{11}(t_k)+u_{21}(t_k))(u_{13}^*(t_k)+u_{23}^*(t_k))\right) |10\rangle_{13}\langle 01|\nonumber\\
&+\frac{1}{2}\mathbb{E}\left((u_{11}^*(t_k)+u_{21}^*(t_k))(u_{13}(t_k)+u_{23}(t_k))\right)
|01\rangle_{13}\langle 10|,
\label{rho13}
\end{align}
as well as the evolution of the state between particles $3$ and $4$
\begin{equation}
\varrho_{34}(t_k)=Tr_{N,N-1,\ldots,5,2,1}(\varrho(t_k)),
\end{equation}
resulting as
\begin{align}
\varrho_{34}(t_k)&=\frac{1}{2}\left[2-\mathbb{E}\left(|u_{13}(t_k)+u_{23}(t_k)|^2\right)
-\mathbb{E}\left(|u_{14}(t_k)+u_{24}(t_k)|^2\right)\right]
|00\rangle_{34}\langle 00|\nonumber\\
&+\frac{1}{2}\mathbb{E}\left(|u_{13}(t_k)+u_{23}(t_k)|^2 \right)|10\rangle_{34}\langle 10|\nonumber\\
&+\frac{1}{2}\mathbb{E}\left(|u_{14}(t_k)+u_{24}(t_k)|^2\right) |01\rangle_{34}\langle 01|\nonumber\\
&+\frac{1}{2}\mathbb{E}\left((u_{13}(t_k)+u_{23}(t_k))(u_{14}^*(t_k)+u_{24}^*(t_k))\right) |10\rangle_{34}\langle 01|\nonumber\\
&+\frac{1}{2}\mathbb{E}\left((u_{13}^*(t_k)+u_{23}^*(t_k))(u_{14}(t_k)+u_{24}(t_k))\right)
|01\rangle_{34}\langle 10|.
\label{rho34}
\end{align}
Then the concurrences ${\cal C}_{13}$ and ${\cal C}_{34}$ can be evaluated in a similar way to Eq.\eqref{cnn}. It results that ${\cal C}_{ij}$ with $i=1\vee 2$, $j\neq1\wedge 2$ is the same of ${\cal C}_{13}$, while $C_{ij}$ with $i\neq1\wedge 2$, $j\neq1\wedge 2$ is the same of ${\cal C}_{34}$.
By numerical evaluation, the concurrences ${\cal C}_{13}$ becomes significantly different from zero (up to $\approx 0.1$) only at short times and then decays.
On the contrary, the concurrence ${\cal C}_{34}$ always results almost negligible. Essentially this means that entanglement (information) is not coherently transferred to any other pair.

\begin{figure}[htb]
\begin{center}
  \includegraphics[scale=0.2]{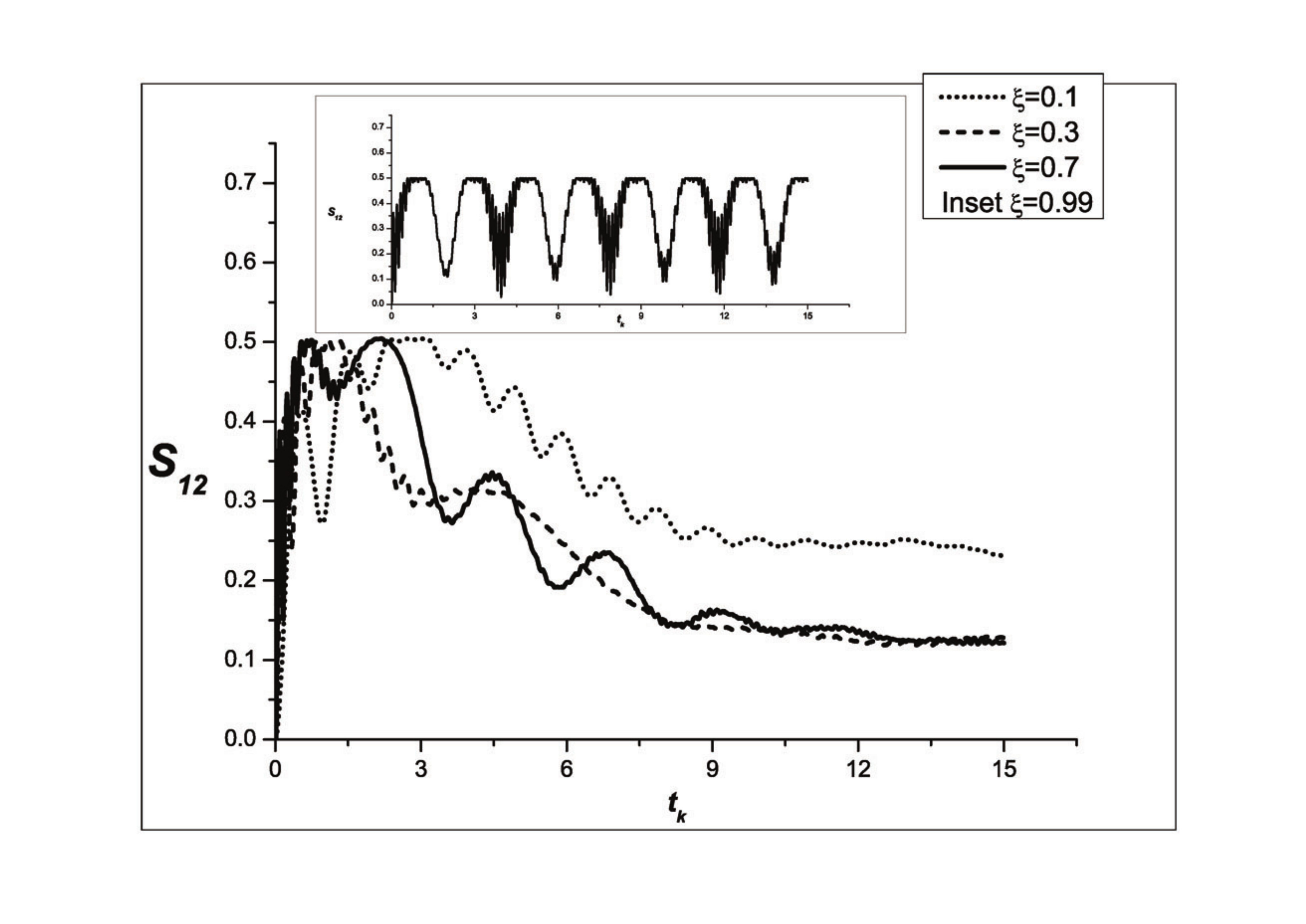}
  \caption{ The linear entropy ${\cal S}_{12}^L(t_k)$ versus (dimensionless) time $t_k$ for $N=32$ and $\Delta t=0.015$.
  }
  \label{fff}
\end{center}
\end{figure}

We also evaluate the linearized von Neumman entropy for the spin particles $1$ and $2$ initially in the pure
state of Eq.\eqref{inistateent} as
\begin{align}
{\cal S}_{12}^L(t_k)=1-Tr\left[\varrho_{12}(t_k)\right]^2.
\end{align}
It ranges from zero for pure state to $3/4$ for a maximally mixed state $\mathbb{I}_{12}/4$ (being $\mathbb{I}_{12}$
 the identity operateor for the qubit $1$ and $2$). Using Eq.\eqref{rho12} we obtain
\begin{align}
{\cal P}_{12}(t_k)&=
\mathbb{E}\left(|u_{11}(t_k)+u_{21}(t_k)|^2\right)
+ \mathbb{E}\left(|u_{12}(t_k)+u_{22}(t_k)|^2\right)\nonumber\\
&-\frac{1}{2}\mathbb{E}\left(|u_{11}(t_k)+u_{21}(t_k)|^2\right)  \mathbb{E}\left(|u_{12}(t_k)+u_{22}(t_k)|^2\right)\nonumber\\
&-\frac{1}{2} \left[\mathbb{E}\left(|u_{11}(t_k)+u_{21}(t_k)|^2\right)\right]^2
-\frac{1}{2} \left[\mathbb{E}\left(|u_{12}(t_k)+u_{22}(t_k)|^2\right)\right]^2\nonumber\\
&-\frac{1}{2} \left| \mathbb{E}\left((u_{11}(t_k)+u_{21}(t_k))(u_{12}^*(t_k)+u_{22}^*(t_k))
\right)\right|^2.
\end{align}
We show in Fig.\ref{fff} the linear entropy ${\cal S}_{12}^L(t_k)$ versus time $t_k$, for a network of $32$ spin particles.
It has a qualitative behavior similar to that of Fig.\ref{ccc}, namely there is a bump (with superimposed small oscillations) occurring sooner for higher values of $\xi$. Also, analogously  to Fig.\ref{ccc} by increasing $\xi$ close to 1 we get an oscillatory behavior with minima approaching zero.
This means that the state $\varrho_{12}$ is periodically purified. Furthermore, in this case the oscillations of the entropy follow those of concurrency ${\cal C}_{12}$ (in contrast to Fig.\ref{ccc} and average fidelity of Fig.\ref{aaa}).
In fact, as a consequence of the symmetry  of the initial state in Eq.\eqref{inistateent} and of the Hamiltonian in Eq.\eqref{H} with respect to the qubit 1 and 2 exchange, each time the state $\varrho_{12}$ is purified it must be
$(|01\rangle+|10\rangle)/\sqrt{2}$, hence ${\cal C}_{12}$ reaches a maximum.

\section{Conclusion}

We have studied information dynamics in a quantum network with underlying graphs where
edges have been added in a random way to mimic unwanted interactions.
We employed the Gilbert model of random graphs characterized by a weighting parameter $\xi$ \cite{Gilbert}.
We have found that by increasing it the dynamics of relevant quantities like fidelity, entropy or concurrence, gradually transforms from damped to damped oscillatory and finally to purely oscillatory.
Actually, the larger is the network size, the wider is the range of $\xi$ from zero for which we have a dissipatory behavior.

It is worth remarking that the weighting parameter $\xi$ could be regarded as a fictitious temperature. In fact
we expect that at low temperature a small number of spin particles interact, i.e. few edges will be present in the graph, while at high temperature a large number of spin particles interact and consequently the underlying graph tends to become fully connected. Then, the presented model encompasses a ``thermal" model for which
$p(n)\propto \exp[-\frac{n}{T}]$,
 with $T$ the temperature, instead of $p(n)\propto \xi^n(1-\xi)^{n_{max}-1}$ in Eq.\eqref{pn}.
In fact with the thermal distribution there is no way to favor the more connected graphs with respect to the less connected ones and we could only obtain the same qualitative results obtainable in the presented model for $\xi<0.5$.

For future one could consider other models of random graphs besides the Gilbert's one.
For instance, exponential random graphs models \cite{West} describe a general probability distribution of graphs
on $N$ nodes given a set of network statistics and various parameters associated to them.
Moreover, there are models of random graphs that figure out ``hubs" that can be interpreted as sinks for information.

The proposed model and related results might be relevant for physical implementation of quantum networks
with various mesoscopic systems, like photonic crystals~\cite{jeremy}, ion traps~\cite{nobel},
superconducting circuits~\cite{franco}, and
planar arrays of trapped electrons used for quantum information processing \cite{penning}.
Furthermore, they could be also useful for theories affording quantum gravity by means of random graphs \cite{Sor}.


\end{document}